# Deep Denoising Method for Side Scan Sonar Images without High-quality Reference Data

Xiaoteng Zhou
School of Ocean Engineering
Harbin Institute of Technology
Weihai, China
zhouxiaoteng@stu.hit.edu.cn

Changli Yu*
School of Ocean Engineering
Harbin Institute of Technology
Weihai, China
* Corresponding author: yuchangli@hitwh.edu.cn

Xin Yuan
School of Ocean Engineering
Harbin Institute of Technology
Weihai, China
xin.yuan@upm.es

Citong Luo
School of Ocean Engineering
Harbin Institute of Technology
Weihai, China
luocitong@gmail.com

*Abstract*—Subsea images measured by the side scan sonars (SSSs) are necessary visual data in the process of deep-sea exploration by using the autonomous underwater vehicles (AUVs). They could vividly reflect the topography of the seabed, but usually accompanied by complex and severe noise. This paper proposes a deep denoising method for SSS images without high-quality reference data, which uses one single noise SSS image to perform self-supervised denoising. Compared with the classical artificially designed filters, the deep denoising method shows obvious advantages. The denoising experiments are performed on the real seabed SSS images, and the results demonstrate that our proposed method could effectively reduce the noise on the SSS image while minimizing the image quality and detail loss.

*Keywords-AUVs; sonar image; image denoising; deep learning;*

## I. INTRODUCTION

In recent years, human exploration of marine resources has gradually migrated to deeper waters. Resources such as oil, natural gas and rare minerals in the ocean will become the main dependence for human survival and industrial development in the future. SSSs are effective equipment carried on AUVs that can help people understand the seabed terrain and geomorphic information. They are not affected by the seabed turbidity, and have a long operating distance and strong anti-interference ability. The SSS images could vividly reflect submarine targets and sediments, so as to help human beings fully understand the ocean environment. Recently, more and more researchers begin to use SSS images for some semantic tasks in computer vision, such as target segmentation, recognition and so on [1,2].

The main reasons that affect the quality of SSS images are image distortion and image noise. Among them, the probability of image distortion is lower and the processing ideas are clear. However, the noise always exists in SSS images and have complex components. At present, there is no mature solution to fully solve SSS image noise problem. The acoustic images and ordinary optical images are quite different in terms of structural features and noise distribution [3]. Echo noise will cause uneven grayscale distribution of SSS images and affect the research progress based on image semantics [4-6]. This noise source signal is mainly formed by the superposition of the acoustic pulse scattered by the scatterers in the sea floor. The noise components of underwater SSS images are complex and superimposed on each other, which are mainly divided into environmental noise, reverberation noise and self-noise. The rougher the terrain, the more obvious the reverberation noise.

The noise on the SSS image is complex and difficult to predict. At present, many researchers use probability statistics to determine their distribution characteristics. A large number of statistical experiments found that the main types of noise on the SSS images are: Gaussian noise, pepper-salt noise and speckle noise, among which speckle noise has the most obvious influences, and the nonlinear gray difference problem caused by it directly affects the application tasks of SSS images in the later stage [7]. Traditional SSS images denoising methods usually assume that the noise model is Gaussian noise or speckle noise. However, in practical underwater applications, the noise on SSS images is usually some complex combinations of many kinds of echo noise. Therefore, the traditional methods based on spatial domain or nonlocal filtering are not ideal for SSS images denoising, because they usually could only deal with a single kind of noise, and need to estimate the noise distribution feature accurately in advance. Additionally, considering the limitation and cost of AUVs acquiring high-quality and clear underwater SSS images, the denoising task of SSS images is facing great challenges.

The main framework of the paper is as follows. Section II reviews the traditional and emerging image denoising methods. Section III presents the detailed methodology of our work. Section IV demonstrates the SSS images for experiment and the experiment details. Section V verifies the algorithm and compares the denoising results. Section VI makes the conclusion and outlook.

## II. RELATED WORK

The traditional methods of processing SSS images mainly refer to the processing ideas of radar images and medical ultrasound images, which are divided into denoising methods based on spatial domain and transform domain, non-local filtering and variational regularization [7]. The SSS denoising methods attempted by some researchers are also based on the above ideas [8-10]. These traditional methods usually require accurate noise estimation to have better effects. They could hardly handle complex noise and do not gengeralize well enough. Additionally, they are easy to destroy the details and texture, which are not optimistic for SSS images processing.

In recent years, with the widespread applications of deep learning technology in the field of image processing, denoising methods based on convolutional neural networks (CNNs) have received more and more attention. These methods have achieved satisfactory results in the denoising research of medical imaging, remote sensing and biometric recognition [11,12]. These methods adopt the data-driven way, use various datasets to train models, and then evaluate and filter image noise. The entire denoising process is close to end-to-end and does not require complex method designs and combinations. Denoising methods based on deep learning are mainly divided into three categories. The first is a supervised method, which uses noisy and clean image pairs of external data sets for training [13]. The limitation of this method is the datasets which require high quality and quantity. Under normal circumstances, it is very difficult and costly to collect enough clean and noisy datasets; the second category is to use only noisy images in external data for training, and no clean data is needed [14]. However, this method requires that the expected processed image has a high correlation with the image used for training. Additionally, in the fields of biomedicine and underwater detection, it is difficult to obtain different noise image pairs of the same target scene. The third category is the way that only one noise image should be used to train and denoise [15], which greatly improves the flexibility.

In the process of AUVs using SSS devices for large-area underwater detection, the same target area is usually detected only once, and the reverberation noise of different seabed positions is different. It is usually difficult to obtain clean or multiple noisy image pairs in the same target area. The proposed training idea in [15] brings a new opportunity to the SSS images denoising task with complex noise components and scarce image data. Based on its framework, this paper designs an SSS image deep denoising method without high-quality reference data. The purpose is to use a self-supervised method to minimize the requirements and assumptions for the input image data to improve the generalization performances of the method, and to introduce a new solution for the denoising task of underwater SSS images.

The main pipeline of our proposal is depicted in Fig. 1. Firstly, input one SSS image and then the regions of interest (ROIs) are extracted and useless regions are eliminated. Next, to classify the sediments into several types, and the parameters of the denoising model are adjusted according to the sediment types. Finally, the quality of the denoised image is evaluated and the best one is output.

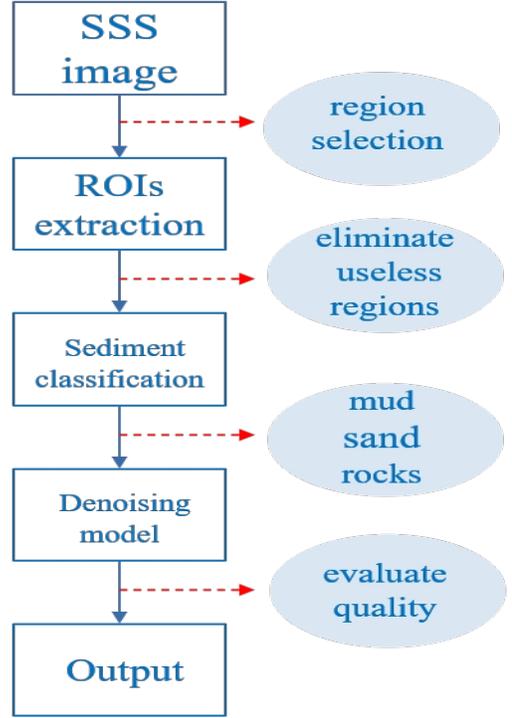

Fig. 1. The main workflow of SSS image denoising.

## III. METHOD

The model design adopts the U-Net architecture in the form of encoder and decoder, which is a self-supervised way. Firstly, one SSS image with complex noise is input and mapped by the encoder to the layer with partial convolution (PConv). The first five encoder blocks are composed of PConv, LReLU and Max pooling layers, and the sixth one has no pooling layer [15]. The decoder part has five decoder blocks to map the image output by the encoder back to the same size as the original image.

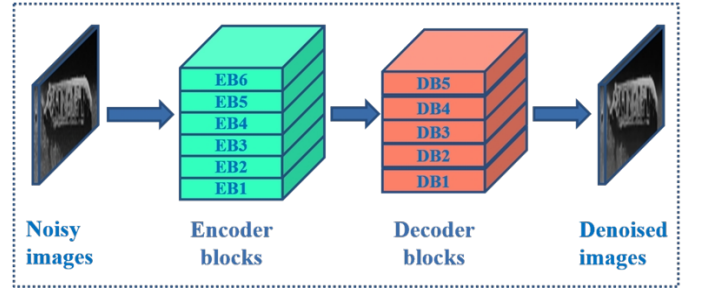

Fig. 2. The main architecture of denoising model.

### A. Image Sampling

An input SSS image is sampled by probability $p$ and Bernoulli sampling rules to obtain a series of noise images, which are used for network training. At the same time, local convolution is used to regularize the pixels of the sampling points twice to improve the performances. Sampling operation and dropout strategy are adopted to deal with the problem of over fitting caused by training the network with a small amount of image data.

## B. Loss Function Design

The loss function is usually used to measure the error between the predicted value and the real value, but in the field of underwater vision research, the real value (clean image) is difficult to obtain, so the variable $\bar{R}$ obtained from Bernoulli sampling needs to be used to replace the real value.

Suppose a binary Bernoulli matrix $S$ composed of 0 or 1, and calculate the sampling result $R$ by dot product with the original image $y$. In addition, the original image and $(1-S)$ dot product operation is used to obtain $\bar{R}$ which is completely opposite to the sampling result. The training network adopts a self-supervised learning model. The single SSS image with noise is input and Bernoulli sampled to obtain multiple similar image pairs, as in

$$\{(R, \bar{R})\}_{m=1}^{M} \tag{1}$$

$$R = S \odot y \tag{2}$$

$$\bar{R} = (1-S) \odot y \tag{3}$$

where $M$ represents $M$ turns of Bernoulli sampling on the input original image $y$ containing noise, $S$ represents the array of Bernoulli distribution, and $S$ is multiplied by the noise image $y$ to obtain the sampled image.

In the training and testing stages, dropout strategy is adopted in order to further reduce the problem of overfitting caused by single image training. The sampled image pair is input for each training, and the loss function is calculated from the image pair as follows:

$$\min_{\theta} \sum_{m=1}^{M} \left\| F_{\theta}(R_m) - (\bar{R}_m) \right\|_{S_m}^{2} \tag{4}$$

where $F_{\theta}$ represents the output of this network, to calculate the $L_2$ loss of each image pair, and then add the $L_2$ loss of $M$ image pairs.

## C. SSS Image Denoising Strategy

In the process of noise reduction (the test stage after model training), Bernoulli sampling is carried out on the specific layer of the trained model. If $N$ Bernoulli samples are taken, it is equivalent to forming $N$ different neural network models to predict the input image. After Bernoulli sampling, the noise image $y$ is successively input into $N$ new neural networks to obtain the prediction output. The final reconstructed image is equal to $N$ prediction outputs. For multiple different output images, average them to obtain desired result $X$, as follows:

$$X = \frac{1}{N} \sum_{n=1}^{N} \hat{X}_n = \frac{1}{N} \sum_{n=1}^{N} F_{\theta}(S_{M+n} \odot y) \tag{5}$$

## IV. EXPERIMENTAL SETTING

### A. Test SSS Images Setting

In this paper, four real SSS images of shipwrecks with complex noise are introduced and tested. The shipwreck 1 in Fig. 3 is from [16] and the shipwrecks 2 to 4 are from [17].

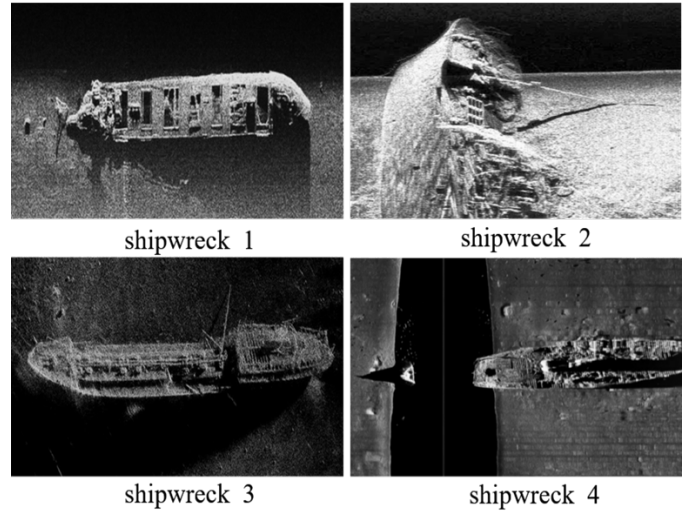

shipwreck 1      shipwreck 2

shipwreck 3      shipwreck 4

Fig. 3. The SSS images for denoising experiment.

### B. Comparison Methods Setting

In order to better display the results of SSS image denoising, we introduce four traditional filters, namely Mean filter, Median filter, Bilateral filter and Wiener filter to carry out comparative experiments.

*1) Mean Filter:* It is a kind of linear filter, and the core is neighborhood average method. Mean filter is the most commonly used approach in image processing, which could eliminate high frequency signals well, so it is helpful to eliminate sharp noise and achieve image smoothing.

*2) Median Filter:* It is a kind of non-linear filter. It has a better effect when dealing with impulse noise and pepper-salt noise, and it could retain more image edge information.

*3) Bilateral Filter:* The filter consists of two functions, one is related to the distance of geometric space; the other is related to pixel interpolation. Bilateral filter is a Gaussian filter function based on spatial distribution, which has one more Gaussian variance than ordinary Gaussian filter, so as to achieve better edge preservation effect.

*4) Wiener Filter:* It is an adaptive filter which could effectively deal with additive noise, and the performances of the Wiener filter are very good under the minimum mean square error optimal criterion.

### C. Test Environment Details

All experiments are implemented under a 64-bit Windows 10 operating system with an Intel Core i7-9700 3.00 GHz processor, 16 GB of physical memory and one NVIDIA GeForce RTX2070s graphics card.

## V. Denoising Results and Evaluation

### A. Evaluation Indexes

For the SSS image preprocessing tasks, due to the poor imaging effect, special attention should be paid to the preservation of image edge and other details in the process of denoising. We introduce peak signal-to-noise ratio (PSNR), structural similarity index matrix (SSIM), flowing index (FI) and edge preservation index (EPI) to quantitatively analyze the results of denoising.

*1) Peak signal-to-noise ratio (PSNR):* It is a similarity evaluation index, which evaluates the image on the basis of mean square error (MSE). The larger the value, the better the denoising effect. The calculation formula is as follows:

$$MSE = \frac{1}{mn}\sum_{i=0}^{m-1}\sum_{j=0}^{n-1}\|I(i,j)-K(i,j)\|^2 \quad (6)$$

$$PSNR = 10 \cdot \log_{10}\left(\frac{(2^n-1)^2}{MSE}\right) = 20 \cdot \log\left(\frac{MAX_I}{\sqrt{MSE}}\right) \quad (7)$$

where $MSE$ is calculated from the input monochrome images $I$ and $K$ of $m \times n$, and $MAX_I$ represents the maximum value of image $I$ point color.

*2) Structural similarity index matrix (SSIM):* Compared with PSNR, SSIM's measurement of image quality is closer to the judgment of human eyes, and it is a good evaluation index when measuring the effect of image distortion. The specific calculation in signal $x$ and signal $y$ is as follows:

$$SSIM(x,y) = [l(x,y)]^\alpha [c(x,y)]^\beta [s(x,y)]^\gamma \quad (8)$$

where $l(x,y)$ compares the brightness, $c(x,y)$ compares the contrast, and $s(x,y)$ compares the structure.

*3) Flowing index (FI):* It refers to the ratio of the mean and standard deviation of pixels in the denoised image. It is usually used to describe the smoothing ability of the filter to noise. Calculate formula is as follows:

$$FI = \frac{M}{SD} \quad (9)$$

*4) Edge preservation index (EPI):* This index is used to measure the edge preservation ability of the image after denoising. The larger the EPI value, the stronger the edge preserving ability of denoising algorithm. It is computed by

$$E = \frac{\sum_{i=1}^{m}|G_{R1}-G_{R2}|_{denoised}}{\sum_{i=1}^{m}|G_{R1}-G_{R2}|_{rawimage}} \quad (10)$$

where $m$ is the number of image pixels, $G_{R1}$ and $G_{R2}$ are the gray values of adjacent pixels in the four directions.

The denoising effects are shown in Fig. 4 and Fig. 5, and the denoising evaluation results are shown in Tables I to IV.

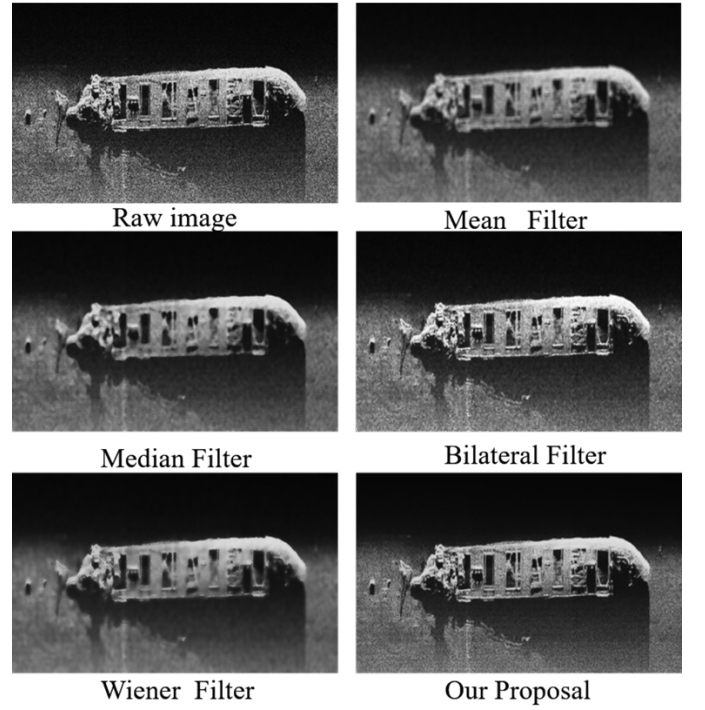

Fig. 4. Overall image denoising effect.

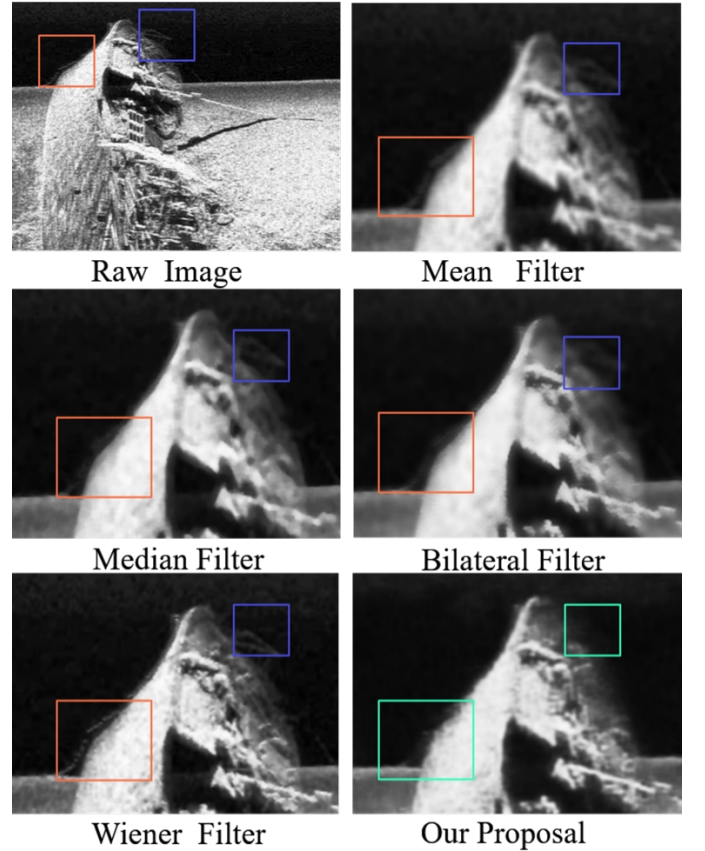

Fig. 5. Schematic diagram of removing stubborn noise.

TABLE I.  EVALUATION RESULTS OF SHIPWRECK 1 DENOISING

| Denoising Method | Evaluation Indexes | | | |
|---|---|---|---|---|
| | PSNR | SSIM | FI | EPI |
| Mean Filter | 21.7271 | 0.5596 | 0.7922 | 0.2054 |
| Median Filter | 21.9978 | 0.5752 | 0.8286 | 0.2310 |
| Bilateral Filter | 23.7420 | 0.6347 | 0.8044 | 0.2338 |
| Wiener Filter | 25.5949 | 0.8366 | 0.8472 | 0.5032 |
| Our Proposal | **28.9961** | **0.8733** | **0.8495** | **0.5309** |

TABLE II.  EVALUATION RESULTS OF SHIPWRECK 2 DENOISING

| Denoising Method | Evaluation Indexes | | | |
|---|---|---|---|---|
| | PSNR | SSIM | FI | EPI |
| Mean Filter | 21.0347 | 0.5587 | 0.6457 | 0.2745 |
| Median Filter | 21.6229 | 0.5803 | 0.6609 | 0.3039 |
| Bilateral Filter | 22.7597 | 0.6086 | 0.6485 | 0.2710 |
| Wiener Filter | 27.1572 | 0.8468 | 0.6664 | 0.5486 |
| Our Proposal | **28.1888** | **0.8688** | **0.6673** | **0.5734** |

TABLE III.  EVALUATION RESULTS OF SHIPWRECK 3 DENOISING

| Denoising Method | Evaluation Indexes | | | |
|---|---|---|---|---|
| | PSNR | SSIM | FI | EPI |
| Mean Filter | 22.7214 | 0.5413 | **0.9551** | 0.1868 |
| Median Filter | 22.7224 | 0.5459 | 0.8999 | 0.2114 |
| Bilateral Filter | 23.7757 | 0.5899 | 0.9545 | 0.1917 |
| Wiener Filter | 28.0258 | 0.8452 | 0.8754 | 0.5583 |
| Our Proposal | **29.1040** | **0.8974** | 0.8862 | **0.5726** |

TABLE IV.  EVALUATION RESULTS OF SHIPWRECK 4 DENOISING

| Denoising Method | Evaluation Indexes | | | |
|---|---|---|---|---|
| | PSNR | SSIM | FI | EPI |
| Mean Filter | 25.2407 | 0.7431 | 0.7754 | 0.3432 |
| Median Filter | 25.4969 | 0.7602 | 0.8106 | 0.3638 |
| Bilateral Filter | 27.3195 | 0.7379 | 0.7826 | 0.3927 |
| Wiener Filter | 34.5559 | 0.9103 | 0.8263 | 0.6733 |
| Our Proposal | **49.4415** | **0.9979** | **0.8391** | **0.9995** |

The evaluation results show that compared with the traditional denoising methods, our proposal could effectively reduce the noise, and the processed SSS images have better smoothness and the edge preserving ability.

## VI. CONCLUSION

In order to solve the limitaion that it is difficult to obtain high-quality SSS reference images for denoising, the deep denoising approach using a single SSS image is proposed. Experiments show that our proposal could effectively denoise and preserve more details. Our proposal has higher flexibility and better generalization performances which could be connected with the subsequent semantic task models based on deep learning to achieve the purpose of end-to-end utilization of SSS images. The future work will mainly establish SSS noise prediction model for various sonar frequencies, seabed geology and detection distance, and then reasonably adjust the denoising parameters to achieve better filtering effect.

ACKNOWLEDGMENT

The research was supported by the Chinese Shandong Provincial Key Research and Development Plan, under Grant No. 2019GHZ011. At the same time, thanks to the sonar image data support provided by SHIPWRECK WORLD and IMGUR.